**Bulk superconductivity and non-trivial band topology analysis of $Pb_2Pd$**

M.M. Sharma[1,2], N.K. Karn[1,2], Poonam Rani[3], R.N. Bhowmik[4], V.P.S. Awana[1,2]

*[1]National Physical Laboratory (CSIR), Dr. K. S. Krishnan Road, New Delhi 110012, India.*
*[2]Academy of Scientific and Innovative Research, Ghaziabad, U.P. 201002, India*
*[3]Materials Science Division, Inter-University Accelerator Centre, New Delhi-110067, India*
*[4]Department of Physics, Pondicherry University, R. Venkataraman Nagar, Kalapet, Pondicherry-605014, India*

**Abstract:**

In this article, we report single crystal growth of superconducting binary compound $Pb_2Pd$. The crystal is well characterized through X-Ray Diffraction (XRD), Selected Area electron diffraction (SAED), Transmission Electron Microscopy (TEM), Field emission Scanning electron Microscopy (FESEM) and X-Ray Photoelectron spectroscopy (XPS). The bulk superconducting nature of the synthesized crystal is determined through AC susceptibility and magneto-heat capacity measurements. The specific heat jump at superconducting transition suggests $Pb_2Pd$ to be a moderately coupled s-wave superconductor. The topological non-trivial character of $Pb_2Pd$ is evidenced through bulk electronic band structure and Z2 invariants, which are calculated under the protocols of Density Functional Theory (DFT). Surface states spectrum of $Pb_2Pd$ is also studied, which further claims $Pb_2Pd$ to have topological non-trivial band structure.

*Corresponding Author

Dr. V. P. S. Awana: E-mail: awana@nplindia.org

Ph. +91-11-45609357, Fax-+91-11-45609310

Homepage: awanavps.webs.com

## Introduction:

The discovery of topological insulators (TIs) has renewed the field of quantum condensed matter. Since then, a flurry of research has been carried out on the materials with symmetry protected topological states, which are characterized by nontrivial topological invariants viz. Z2, Z4 and the Chern numbers [1-4]. Topological materials are classified in different categories based on their band structure and the symmetry such as topological insulators, topological crystalline insulators (TCIs) and topological semimetals (TSMs) etc. [1,5]. Moreover, TSMs are further sub-classified as Dirac Semimetal (DSM), Weyl Semimetal (WSM) and Nodal line semimetal (NLSM) [5-7]. Among them, DSMs has fourfold degenerate band crossings and WSMs have pairs of Weyl nodes, which are generated due to broken Time Reversal symmetry (TRS) and Inversion symmetry [6]. Apart from these three categories semimetals are classified into two other categories such high-symmetry-point semimetal (HSPSM) and high-symmetry-line semimetal (HSLSM) [8]. These materials show band degeneracy at high symmetry points in the bulk electronic band structure calculated without considering Spin Orbit Coupling (SOC) [8].

TSMs with bulk superconducting properties are emerged as the materials of interest as these are regarded as the possible candidates to realize topological superconductivity [9-19]. Topological superconductors are considered as the key materials to host Majorana fermions [20,21], while some of the p wave superconductors also show the existence of Majorana bound states in their vortices [22,23]. Some of the TSMs viz. $PdTe_2$ [9], $PbTaSe_2$ [10], $SnTaS_2$ [11], NbC [12]. $AuSn_4$ [13], $Au_2Pb$ [14] show bulk superconductivity with topological non-trivial surface states. Apart from these materials some binary compounds viz. BiPd, β-$Bi_2Pd$ and $Sb_2Pd$ are found to show topological states with bulk superconductivity [15-17]. Theoretically, thin films of β-$Bi_2Pd$ are supposed to show topological superconductivity and Majorana zero modes [18,19], while this feature lacks experimental confirmation. $Pb_2Pd$ is another similar compound, and its superconducting nature has been known since 1962 [24]. $Pb_2Pd$ is not much studied in the context of its topological properties as there is only one theoretical report available in which $Pb_2Pd$ is considered as a TSM [8]. Interestingly, superconducting properties of $Pb_2Pd$ also have not been explored to that extent. There are only two reports available in literature about superconducting properties of $Pb_2Pd$ [25,26]. The first report was on the effect of Bi substitution in $Pb_2Pd$ [25] and the other one is a very recent report showing type-I superconductivity in the same [26]. It is important to study $Pb_2Pd$ in the context of superconductivity and topological properties as it is supposed to be a superconducting material with topological non-trivial band structure [8,26].

In this article, we report the synthesis of $Pb_2Pd$ single crystal using solid state reaction route, this sample is well characterized through various characterization techniques. Bulk superconductivity is the key issue for superconducting topological materials, and the same with $T_c^{onset}$ at 2.86K is confirmed in the synthesized $Pb_2Pd$ crystal through AC susceptibility and magneto-heat capacity measurements. This is the first report showing magneto-heat capacity measurements for $Pb_2Pd$ crystal. Further, the topological properties are studied through DFT calculations and theoretical simulation of topological invariants. The Z2 invariants of $Pb_2Pd$ shows the presence of strong topology, which is further verified by studying the surface state spectrum. It is worth mentioning here that this is the first report on such topological invariants for $Pb_2Pd$.

**Experimental:**

Solid state reaction route is used to synthesize the single crystalline sample of $Pb_2Pd$. 4N pure powders of Pb and Pd were mixed in 2:1. The mixture was made homogenous by grinding it inside MBraun Glove Box filled with Argon gas. This homogenous mixture was then palletized and vacuum encapsulated in quartz ampoule under the pressure of $5 \times 10^{-5}$mbar. This ampoule was heated to 900°C at 120°C/h for 24 hours. At this higher temperature Pb itself acts as flux and helps Pd to melt, despite of its higher melting point. The hold time of 24 hours at 900°C is utilized in melting the sample and to homogenize the melt. Then the sample is slowly cooled to 460°C at 2°C/h and then kept at this temperature for 24 hours. After that, the sample is quenched at 460°C in water. The step of quenching at 460°C is very crucial as the phase formation temperature of $Pb_2Pd$ and PbPd is very close and the both phases have simultaneous existence at around 445°C [27]. The schematic of heat treatment is shown in Fig. 1.

Rigaku Miniflex II table top XRD equipped with Cu-$K_\alpha$ (1.5406Å) was used for phase purity characterization of synthesized $Pb_2Pd$ single crystal. Surface morphology and elemental composition were analyzed using MIRA II LMH TESCAN made Field Emission Scanning Electron Microscope (FESEM) equipped Energy Dispersive X-Ray (EDX) detector. JEOL/JEM-F200 with One view CMOS camera (4K*4K) Transmission Electron Microscope (TEM) is used to get Side Angle Electron Diffration (SAED) pattern and to visualize atomic planes. XPS spectra was recorded by using X-ray Photoelectron Spectrometer (XPS) (Model:

K-Alpha-KAN9954133, Thermo Scientific) with monochromated and micro-focused Al Kα radiation (1486.6 eV). The spectrometer has been designed with a dual-beam flood source to provide a charge compensation option and calibration is confirmed by the position of C 1 s line at the binding energy 284.8 eV. The working pressure was maintained < $6 \times 10^{-8}$ mbar. The scans for Pb 4f and Pd 3d were recorded with the step size of 0.1eV. AC susceptibility and Heat capacity measurements were carried out using Quantum Design Physical Property Measurement System (QD PPMS) in temperature range from 5K to 2K. FullProf software and VESTA software were used for Rietveld refinement and the unit cell construction respectively.

To study and understand the topological properties of synthesized $Pb_2Pd$ crystal, the first principle simulations have been carried out and Z2-invariants are also calculated to categorize the topology present in the system. For this, Density Functional Theory (DFT) based on first principle calculations are executed in QUANTUM ESPRESSO [28,29] to obtain the bulk electronic band structure and density of states (DOS). For the calculations, atomic positions of primitive unit cell of $Pb_2Pd$ are considered. The Perdew-Burke-Ernzerhof (PBE) generalized gradient approximation (GGA) was used to account for the electronic exchange and correlation. The wave functions were expanded in a plane wave with Gaussian smearing of the width 0.01 on a Monkhorst-Pack k-grid of 8×8×8. The electronic bands were calculated with and without the inclusion of SOC. For band structure calculated without-SOC, we used Standard Solid-State Pseudopotential (SSSP) library. For with-SOC band calculations, we used the PSEUDODOJO library, which incorporates full relativistic approximations. For the convergence of self-consistent calculation cutoff is $1.2\times10^{-9}$ Ry and charge cut-off of 320 Ry and wave function cut-off 45 Ry was used. For calculation of Z2 invariants, DFT generated Bloch wave function is wannierized using the *WANNIER90* software [30]. The WANNIER90 uses the Wannier functions which are the representation of Block wave functions in real space [31,32]. Out of 108 Bloch wave-functions, 44 are wannierized and disentanglement calculation converges with threshold limit $10^{-10}$. The band structure is reproduced in the energy range $E_F \pm 2$eV by selecting projectors p orbitals of Pb and d orbitals of Pd atoms with the whole Brillouin Zone sampled on a dense 8×8×8 K-mesh. The band structure is again obtained using the Maximally localized Wannier Function (MLWFs) and it is verified by the band structure generated through first principle method. An effective tight-binding (TB) model for $Pb_2Pd$ crystal was obtained using MLWFs after wannierization of Block wavefunction. This effective TB model is further processed and applied in WANNIER-TOOLS [33]. The whole Brillouin Zone is sampled on a 8×8×8 K-mesh, for all calculations. The states of Z2-invariants are determined by the evolution of Wannier charge centers in Brillouin zone planes which are sampled on much dense grid of 61×61×21. Further, surface spectral function is calculated using the iterative surface Green's function method [34,35] along the plane (011).

**Result & Discussion:**

Fig. 2 represents the Rietveld refinement results of powder XRD (PXRD) pattern of synthesized $Pb_2Pd$ single crystal. The synthesized $Pb_2Pd$ crystal crystallizes in tetragonal structure with I4/mcm space group. All peaks of PXRD pattern are well fitted with the phase parameters and confirms that the synthesized crystal is phase pure. The quality of refinement is determined through $\chi^2$ value which is found to be 4.95, which is reasonably good. The constituent elements viz. Pb and Pd occupies the atomic positions viz. (0.165,0.665,0) and (0,0,0.25) in a tetragonal unit cell. The Rietveld refined lattice parameters are $a = b = 6.863(5)Å$ & $c = 5.840(5)Å$ and $\alpha = \beta = \gamma = 90°$. All these parameters are in good agreement with previous report on the same compound [26]. The unit cell parameters along with the refinement parameters are listed in table-1. The inset of Fig. 2 represents the VESTA drawn unit cell of synthesized $Pb_2Pd$ crystal. Crystallographic information file (CIF) generated from Rietveld refinement is used to draw the unit cell. The synthesized crystal is shown to have a body-centered tetragonal unit cell with Pd atoms at the body center.

Fig. 3(a) shows SAED pattern of synthesized $Pb_2Pd$ single crystal. SEAD pattern was taken on powdered sample. Spots from (002), (004) and (220) are encircled in Fig. 3(a). Crystalline nature of synthesized $Pb_2Pd$ crystal is evident from TEM image, shown in Fig. 3(b). Fig. 3(b) shows that the atomic planes are stacked in a regular manner along a specific growth direction. The calculated interplanar spacing between the atomic planes as shown in TEM image is found to be 2.94Å, which corresponds to the interplanar spacing of (002) planes. Stacking of atomic planes in (002) plane direction as revealed by TEM image, suggests that the crystal of $Pb_2Pd$ is grown in (00l) direction, it is also consistent with previous report [26]. To get more insight about crystalline nature of synthesized $Pb_2Pd$, surface morphology is visualized through FESEM image. Fig. 3(c) shows the FESEM image of the synthesized $Pb_2Pd$ single crystal. The FESEM image is showing terrace type morphology which represents layer by layer growth of the sample. No grain is visible in the FESEM image which also confirms the single crystalline nature of the synthesized crystal. Elemental composition is determined through Energy Dispersive X Ray Analysis (EDX) measurements and the results are shown in Fig. 3(d). The EDX spectra shows the presence of constituent elements viz. Pb and Pd. No impurity element is detected in EDX spectra which shows the purity of the synthesized crystal. Elemental composition is shown in inset of Fig. 3(d), both the elements are found to be in near stoichiometric ratio.

Chemical states of the constituent elements of $Pb_2Pd$ are analyzed through XPS spectra. XPS peak of C 1s is taken as reference to calibrate all XPS peaks. Fig. 4(a) and 4(b) show XPS spectra of synthesized $Pb_2Pd$ crystal in Pb 4f and Pd 3d regions respectively. These spectra are fitted with Gaussian distribution function. The XPS spectra in Pb 4f region is deconvoluted in four peaks viz. P1, P2, P3 and P4. Peaks P1 and P3 correspond to spin orbit doublet of Pb 4f viz. Pb $4f_{5/2}$ and Pb $4f_{7/2}$. These peaks occurred due to core levels of $Pb^{2+}$ cations and the corresponding binding energies are found to be 138.55eV and 143.42eV for Pb $4f_{7/2}$ and Pb $4f_{5/2}$ respectively. These values are shifted from the binding energy values for metallic Pb [36], this shift in binding energies is occurred due to bonding with Pd atoms. This suggests strong covalent bonding between Pb and Pd. The remaining peaks viz. P2 and P4 corresponds to $Pb^{2+}$ cations resulting from PbO formation. These peaks occurred due to surface oxidation of the synthesized sample due to air exposure. The peaks of $Pb^{2+}$ ions generated due to PbO formations is less pronounced in comparison to that observed due to $Pb^{2+}$ cations of synthesized $Pb_2Pd$ crystal. The energy separation between the spin orbit doublets of Pb is found to be 4.87eV, which is very close to the standard value (4.86eV) [36]. Fig. 4(b) shows the XPS spectra in Pd 3d regions, which consists of XPS peaks due to spin orbit doublet of Pd 3d core levels viz. Pd $3d_{5/2}$ and Pd $3d_{3/2}$. The peaks of spin orbit doublets viz. Pd $3d_{5/2}$ and Pd $3d_{3/2}$ are observed at 335.85eV and 341.16eV respectively. These values are slightly shifted from the standard value due to bonding of Pd with Pb atoms [36]. The energy separation of these peaks is found to be 5.31eV which is very close to the standard value (5.26eV) [36]. These XPS results suggest that the valency of Pb atoms in $Pb_2Pd$ is +2. All XPS peaks for spin orbit doublet of constituent elements viz. Pb and Pd along with their respective full width at half maxima (FWHM) are listed in table-2.

Fig. 5 shows the results of AC susceptibility measurements carried out at different AC magnetic field viz. 3Oe, 5Oe, 7Oe, 9Oe and 11Oe. During the whole AC magnetization measurement, the frequency of the AC field was set at 333Hz and the background DC field was made stable to 0Oe. Real part of AC magnetic susceptibility ($\chi'$) is shown in the lower plot while the imaginary part ($\chi''$) of the same is shown in upper one. Both the real and imaginary parts of the magnetic susceptibility show presence of bulk superconductivity in the synthesized $Pb_2Pd$ crystal. The onset superconducting transition temperature ($T_c^{onset}$) is found to be 2.9K.

Generally, AC susceptibility measurements give the hint about the granularity of the sample. For a granular superconductor, $T_c$ tends to shift to lower temperature with increasing the amplitude of the AC field. This effect occurs due to the intergranular coupling of superconductors. The shift of $T_c$ with respect to the change in AC field amplitude determines the strength of intergranular coupling [37]. Here, $T_c$ remains constant when the amplitude of the AC field is increased. This suggests that grains are absent due to which intergranular $T_c$ could be evolved. This also determines the crystalline character of the synthesized $Pb_2Pd$ crystal. For quantitative analysis of superconducting properties through AC magnetization measurements, it is important to consider the demagnetization factor of measured sample. Demagnetization factor depends on the shape of the sample, and for a perfectly cylindrical sample its value is accounted to be 1. Here due to crystalline nature of the sample, measurements have been carried out on a mechanically cleaved rectangular flake. For a rectangular superconducting sample, demagnetization factor can be calculated by the following formula as suggested in ref. [38]

$$N = 1 - \frac{1}{1+\frac{q*a}{b}} \quad (1)$$

Where a and b are dimensions of the sample in perpendicular to applied field and parallel to the applied field, and these are 3.1mm and 0.21mm respectively. The other parameter q is shape dependent parameter and for a rectangular sample it is determined by the following formula

$$q = \frac{\pi}{4} + 0.64 tanh\left[0.64\frac{b}{a} ln\left(1.7 + 1.2 * \frac{a}{b}\right)\right] \quad (2)$$

The value of q is found to be 0.866 and the corresponding value of demagnetization factor is found to be N=0.927. Demagnetization factor is used to determine effective AC susceptibility by using the relation $\chi_{eff} = \frac{\chi_m}{1+N*\chi_m}$, where $\chi_m$ is measured AC susceptibility. The synthesized $Pb_2Pd$ crystal is found to show a superconducting volume fraction about 42%, which is quite low as compared to the standard value for a perfect bulk superconductor. This low volume fraction is observed due to surface oxidation of the synthesized $Pb_2Pd$ crystal due to air exposure. $Pb_2Pd$ single crystal is prone to surface oxidation, this is also observed in XPS measurements.

Heat capacity measurements are the most reliable method to determine bulk superconductivity in a superconducting sample. Heat capacity measurements are carried out on synthesized $Pb_2Pd$ crystals at different magnetic fields viz. 0Oe, 20Oe, 30Oe and 50Oe. Heat capacity divided by T i.e., C/T is plotted against the temperature and the same is shown in inset of Fig. 6(a). A clear heat capacity jump is visible with $T_c^{onset}$ at 2.86K. This confirms the presence of bulk superconductivity in the synthesized $Pb_2Pd$ single crystal. Heat capacity of a material is mainly contributed by two terms; first one is electronic term ($C_{el}$) and the other is phonon term ($C_{ph}$). These are described by the following equation

$$\frac{C}{T}=\gamma_n+\beta_n T^2 \text{ or } C=\gamma_n T+\beta_n T^3 \quad (3)$$

In the above equation the first term $\gamma_n T$ represents electronic contribution to heat capacity and the second term $\beta_n T^3$ phonon contribution to heat capacity. The coefficient associated with these terms can be determined by linearly fitting the C/T vs $T^2$ plot with equation (3). The Fig. 6(a) represents a linearly fitted C/T vs $T^2$ plot of synthesized $Pb_2Pd$ single crystal. The coefficient of $C_{el}$ i.e. $\gamma_n$ is known as the Sommerfeld coefficient and it is found to be 5.72 ± 0.34 mJ $mol^{-1}$ $K^{-2}$. The other constant term $\beta_n$ is found to be 3.18 ± 0.02 mJ $mol^{-1}$ $K^{-4}$. The value of $\gamma_n$ is used to determine the density of states at Fermi level [$D_c(E_F)$] using the formula,

$$\gamma_n = \pi^2 k_B^2 D_c(E_F)/3 \quad (4)$$

Here, $k_B$ is Boltzmann constant. The obtained value of $D_c(E_F)$ is 2.35 states $eV^{-1}f.u.^{-1}$. The coefficient associated with phonon contribution term i.e., $\beta_n$ is related to Debye temperature with the following formula

$$\theta_D = \left(\frac{12\pi^4 nR}{5\beta_n}\right)^{1/3} \quad (5)$$

Here R=8.314 J $mol^{-1}$ $K^{-2}$, n=3 (for $Pb_2Pd$). The obtained value of $\theta_D$ is 122 ± 1K. Now the values of $\theta_D$ and $T_c$ are used to determine the electron phonon coupling constant $\lambda_{e\text{-}ph}$ using the McMillan formula [39] given below

$$\lambda_{e-ph} = \frac{1.04 + \mu^* ln(\theta_D/1.45T_C)}{(1 - 0.62\,\mu^*) ln(\theta_D/1.45T_C) - 1.04} \quad (6)$$

Here, the value of $\mu^*$ is taken to be 0.13 as taken in ref. [26] and it is known as the screened repulsive Coulomb potential. $\mu^*$ can be assigned any value between 0 to 0.2 for superconductors with $T_c$ below 20K [39]. The empirical value of $\mu^*$ suggested in ref. 39 is 0.13, which is taken in present study. The similar value of $\mu^*$ is taken for other Pb based heavy intermetallic superconductors [40]. The obtained value of $\lambda_{e\text{-}ph}$ is 0.70 and it is higher than the usual values that are obtained for weakly coupled superconductors. This suggests moderate coupling in synthesized $Pb_2Pd$ single crystal. The obtained value of $\lambda_{e\text{-}ph}$ is verified through theoretical studies in later part of this article.

Now, the electronic heat capacity ($C_{el}$) is calculated by subtracting the phonon term from the total heat capacity. The normalized electronic heat capacity is plotted against $T/T_c$ and is shown in Fig. 6(b). This is used to determine the magnitude of heat capacity jump i.e., $\Delta C_{el}/\gamma_n T_c$, which is found to be 1.83. This value is higher than the BCS weak coupling limit which is 1.43. This also suggests that the synthesized $Pb_2Pd$ is a moderately coupled superconductor, which agrees with the obtained value of $\lambda_{e\text{-}ph}$. Normalized electronic heat capacity also gives the information about the superconductivity of the sample, whether it is a conventional superconductor or an unconventional superconductor. Also, it is used to determine the value of the parameter $\alpha = 2\Delta(0)/k_B T_C$. For this, the normalized specific heat data is fitted with the s wave equation and it is found to be well fitted with the same, showing $Pb_2Pb$ to be a bulk superconductor with conventional s-wave pairing. The obtained value of $\alpha$ from the fitted plot is 5.68, which is also higher than the BCS value for weakly coupled superconductors. The corresponding value of superconductivity energy gap at absolute zero $\Delta(0)$ is found to be 0.70meV. All above discussed parameters suggest that synthesized $Pb_2Pd$ single crystal is a moderately coupled superconductor with conventional s-wave pairing; these results are in agreement with the previous report on the same compound [26].

Fig. 6(c) shows the results of heat capacity vs T measurements at different field viz. 0Oe, 20Oe, 30Oe and 50Oe. In this plot, C/T is plotted against T to visualize the heat capacity jump more clearly. These measurements give the value of $T_c^{onset}$ to be 2.86K, 2.75K, 2.66K and 2.40K at 0Oe, 20Oe, 30Oe and 50Oe respectively. It can be clearly seen that the $T_c$ is decreasing with increasing field. Critical field is plotted against the temperature and fitted with the following equation

$$H_c(T) = H_c(0) * \left(1 - \frac{T^2}{T_C^2}\right) \tag{7}$$

where $T_c$ is taken to be 2.86K. The fitted plot is shown in inset of Fig. 6(c), this fitted plot gives the value of critical field at absolute zero, $H_c(0)$ to be 260Oe, which is comparable to that was obtained in previous report [26]. The value of $H_c(0)$ is verified through magnitude of heat capacity jump. Sommerfeld coefficient and heat capacity jump are related to $H_c(0)$ through the following formula [41,42],

$$\triangle C = \frac{4H_c(0)^2}{\mu_0 T_c} = 1.43\gamma_n T_c \tag{8}$$

here, $\Delta C$ and $\gamma_n$ are taken in per unit volume. The molar volume of $Pb_2Pd$ is found to be $4.14\times10^{-5}m^3$/mole and the magnitude of heat capacity jump, $\frac{\triangle C}{\gamma_n T_c}$ is taken to be 1.83. The obtained value of $H_c(0)$ is found to be 249Oe, which is nearly equal to 260Oe, which is obtained from the fitted plot.

Moreover, the $C_{el}$ is determined at the various applied fields. $C_{el}/T$ is plotted against T and is shown in Fig. 6(d). This gives the value of $\gamma$ at different fields at T=2K. The values of $\gamma$ are normalized with $\gamma_n$ and are plotted against the applied field. This gives the important information regarding the low energy excitations which exists in the proximity the Abrikosov vortex line. In conventional superconductors, these low energy excitations take place inside the vortex cores in normal states having the radius which are proportional to penetration depth ($\xi$). For this, specific heat in superconducting state is proportional to vortex density and linearly depends on magnetic field giving $\gamma(H) \propto H$ [43]. While, in case of superconductors having nodes in energy gap, the DOS are found in the neighborhood of gap nodes. Due to this, the low energy excitations occur outside the vortex core and the specific heat is found to show square root dependence on magnetic field given as $\gamma(H) \propto H^{1/2}$, this is known as Volovik effect [44]. In the present case, $\gamma(H)$ is found to have a linear relationship with the applied field as shown in inset of Fig. 6(d). This also confirms that the observed bulk superconductivity in synthesized $Pb_2Pd$ single crystal is conventional superconductivity.

To determine topological non-trivial character of synthesized $Pb_2Pd$ crystal, bulk electronic band structure and DOS are calculated using DFT protocols. The band structure calculation is performed using the cell parameters obtained from the Rietveld refinement. Both the SOC and without SOC protocols are executed in calculations as these are applied in Quantum Espresso with Perdew-Burke-Ernzerhof (PBE) exchange-correlation functional [28,29]. The k-path which is followed for calculations is determined from the SeeK-path: the k-pathfinder and visualizer [45], which eventually suggests that the k-path to be Z →Γ → M→ X →Γ →N → P →X is an optimized path. Fig. 7(a) depicts this particular k-path, as it is marked in the First Brillouin zone. The DFT calculated bulk electronic band structure is shown as the left-hand side image of Fig. 7(b) while the DOS are shown in the right one. The calculated DOS and bulk electronic band structure indicates that the bands near Fermi level have major contributions from the d-orbitals of Pd and p-orbitals of Pb atoms. There are four bands which cross Fermi level, confirming Metallic/Semi-metallic behavior of the crystal. The Fermi surfaces corresponding to these four bands are calculated using WANNIER TOOLS and plotted in XCRYSDEN [46], the same are shown in Fig. 7(a). The DOS at Fermi level are found to be 2.74 states $eV^{-1}$ per primitive unit cell, and there are two formula units of $Pb_2Pd$ in its primitive unit cell, so there will be 1.37 states $eV^{-1}$ $f.u.^{-1}$. The difference between the theoretical and experimental values of DOS at Fermi level hints towards moderate coupling of electrons and phonons. This value of DOS at fermi level is used to calculate theoretical Sommerfeld coefficient $\gamma_b$, which is found to be 3.34 mJ $mole^{-1}$ $K^{-2}$. Electron phonon coupling constant $\lambda_{e\text{-}ph}$ can also be calculated using the formula $\gamma_n = \gamma_b * (1 + \lambda_{e-ph})$. The obtained value of $\lambda_{e\text{-}ph}$ is found to be 0.71, which is in agreement to the value obtained from Macmillan formula, in which the value of $\mu^*$ was taken to be 0.13. This shows that the theoretical results agree with the experimental one. The experimental and theoretical values of Sommerfeld coefficient viz. $\gamma_n$ and $\gamma_b$ help to determine effective mass ($m^*$) of the system, $m^*$ can be calculated in terms of electronic mass $m_e$ using the relation $\frac{\gamma_n}{\gamma_b} = \frac{m^*}{m_e}$ [47]. The obtained value of $m^*$ is found to be $1.71m_e$. Here an exercise is made to determine the nature of superconductivity in $Pb_2Pd$, as the ref. 26 shows the same to be type-I superconductor while it is shown to be of type-II in a very recent report [48]. Here, nature of superconductivity in synthesized $Pb_2Pd$ crystal is determined by calculating Ginzberg Landau Kappa ($\kappa$) parameter. The $\kappa$ parameter is given as the ratio of superconductivity characteristic lengths viz. London penetration depth i.e. $\lambda(0)$ and BCS coherence length $\xi(0)$, and represented as $\kappa = \frac{\lambda(0)}{\xi(0)}$. The London penetration depth is calculated by using the relation $\lambda(0) = \left(\frac{m^*}{\mu_0 n e^2}\right)^{1/2}$, and its value is found to be 27.36nm. $\xi(0)$ is calculated by using the relation $\xi(0) = \frac{0.18\hbar^2 k_F}{k_B T_c m^*}$. Here $k_F$ is Fermi wave vector and is calculated by considering a spherical Fermi surface. For spherical Fermi surface, Fermi wave vector is given by $k_F=(3n\pi^2)^{1/3}$, where n is carrier concentration. $Pb_2Pd$ provides 8 electrons per unit cell, so total electron density can be calculated by determining the ratio of number of electrons per unit cell and total volume of a unit cell i.e. n=8/V, where V=275.068Å$^3$. The value of n is found to be $2.90\times10^{28}$ $m^{-3}$, and the corresponding value of $k_F$ is found to be 0.95 Å$^{-1}$. This value of Fermi wave vector is used to calculate the BCS coherence length, which is found to be 230.98nm. $\kappa$ parameter is calculated by taking the ratio of both the characteristics lengths and it is found to be 0.12, which is much less than $1/\sqrt{2}$, the cut off value for type-II superconductivity. Our results shows that $Pb_2Pd$ is a type-I superconductor, which is in agreement with muon spin rotation ($\mu$-SR) measurements in ref. 26.

Moreover, bands are found to show line degeneracy which is eventually lifted when SOC parameters are included in the calculations. This indicates the effectiveness of SOC in the studied $Pb_2Pd$ system. Particularly, along the path N $\rightarrow$ P, the bands near fermi level are doubly degenerate forming nodal lines. Interestingly, when SOC parameters are included in calculation, the line degeneracy has been lifted except from a few points where double degeneracy remains intact. The region showing the bands along path N $\rightarrow$ P, is shaded in Fig. 7(b). The zoomed view of the shaded region is shown in Fig. 7(c). Nodal lines are clearly visible in Fig. 7(c) and the line degeneracy is found to be removed when SOC parameters are considered. Fig. 8(a) shows (011) plane in first Brillouin zone, for which gap energy dispersion has been calculated. The gap energy dispersion is shown in Fig. 8(b), in which the image in the left panel shows without SOC dispersion and the right panel shows the same with inclusion of SOC parameters. In without SOC dispersion, closed loop line degeneracy can be seen but when SOC is included, all line degeneracy seems to be lifted. Therefore, $Pb_2Pd$ is not a nodal line semimetal. However, along the high symmetry path $\Gamma \rightarrow$ M and M $\rightarrow$ X$\rightarrow \Gamma$ in the bulk electronic band structure in fig.7(b), we find that there exists a type-II Dirac cone like band structure, where degeneracy is lifted when SOC parameters are included. The presence of Dirac points along the high symmetry lines in with SOC bands suggest that $Pb_2Pd$ can be classified as a high symmetry line semimetal which is also in agreement with the previous report [8], which is a catalogue of all possible topological materials.

Further, we calculated the surface state spectrum using the iterative Green's function implemented in Wannier Tools. The surface card is set to be plane (011), and the calculation was done by taking 101 slices of one reciprocal vector. The plane (011) is shown in figure 8(a), and in that plane, the path taken for surface state spectrum calculation is $\overline{\mathrm{M}} \rightarrow \overline{\mathrm{X}} \rightarrow \overline{\mathrm{M}}$. Figure 8(c) shows the obtained spectrum, where a possible Dirac cone is observable at $\overline{\mathrm{X}}$ point near energy -1.0eV. A similar Dirac cone of type-II is observable very close to fermi energy towards high symmetry point $\overline{\mathrm{M}}$ but not at the high symmetry point. Both the surface state spectrum and bulk electronic band structure indicates that the studied $Pb_2Pd$ crystal is a topological material.

It is important to characterize the topology present in a topological material in terms of a topological invariant. These topological invariants depend on the symmetry preserved by the material. The observed splitting of bands with inclusion of SOC parameters in bulk electronic band structure, suggests that the system respects the TRS and these kinds of topological systems are characterized in terms of Z2-invariants [49]. Here, we follow the Soulyanov-Vanderbilt [49] method of Wannier Charge Centers (WCC), calculated from MLWFs to calculate Z2 invariants. The WCCs are the pseudo charge points which are the locations of extremum probability points distribution of MLWFs, maximum probability point regarded as negative charge and with least (zero) probability point are regarded as positive charge. Exchange of these charge centers in different k-planes define trivial and non-trivial topology. The evolution of WCC take place in the 6-planes in the Brillouin zone, namely $k_1$, $k_2$, $k_3$ = 0 and $k_1$, $k_2$, $k_3$ = 0.5 The calculated Z2 invariants for the $Pb_2Pd$ crystal in the mentioned six planes are

(a) In $k_x = 0.0$ i.e. $k_y$–$k_z$ plane: Z2 = 1.
(b) In $k_x = 0.5$ i.e. $k_y$–$k_z$ plane: Z2 = 0.
(c) In $k_y = 0.0$ i.e. $k_x$–$k_z$ plane: Z2 = 1.
(d) In $k_y = 0.5$ i.e. $k_x$–$k_z$ plane: Z2 = 0.
(e) In $k_z = 0.0$ i.e. $k_x$–$k_y$ plane: Z2 = 1
(f) In $k_z = 0.5$ i.e. $k_x$–$k_y$ plane: Z2 = 0

The topological Z2 index is represented by four parameters as $(\upsilon_0; \upsilon_1\upsilon_2\upsilon_3)$, where $\upsilon_0$ represents the strong index and $\upsilon_1, \upsilon_2, \upsilon_3$ represent the weak index. The weak index is the values of Z2 numbers for $k_i$=0.5 planes (i=x, y and z). The strong index may have some redundancy as the strong index is given by Z2 value for all three planes $k_i$=0(i=x, y and z) and Z2 value may not be the same for all three planes. But here we find that for all three planes, $\upsilon_0 = 1$, indicating a topologically non-trivial state and strong topology present in the system. Here the weak index has no redundancy, and thus Z2 index is $(1; 000)$ for the studied $Pb_2Pd$ system. The same Z2 index is reported for $Bi_2Te_3$, which is considered as a strong topological material [50]. Recently, another group [48] has reported the same non-trivial Z2 index $\upsilon_0 = 1$ using a different method – evaluating the band parity at high symmetric time reversal invariant momenta points Γ and M [51]. Thus, on the basis of theoretical calculations, it can be concluded that $Pb_2Pd$ has strong non-trivial band topology.

**Conclusion:**

Summarily, we have presented detailed analysis of bulk superconducting and non-trivial topological properties of self-flux grown $Pb_2Pd$ single crystal. $Pb_2Pd$ is first time characterized through TEM and XPS techniques in this paper. Pb is found to be $Pb^{2+}$ state in $Pb_2Pd$. The synthesized $Pb_2Pd$ crystal is found to be a bulk superconductor with $T_c^{onset}$ at 2.86K. The calculated parameters viz. electron phonon coupling constant $\lambda_{e\text{-}ph}$, $\Delta C_{el}/\gamma T_c$ and α confirm that $Pb_2Pd$ is a moderately coupled s wave superconductor. Conventional superconductivity is also confirmed through magnetic field dependence of the Sommerfeld coefficient in its superconducting state. The calculated bulk electronic band structure gives the glimpses of band inversion with including SOC parameters and suggest $Pb_2Pd$ to be a possible topological material with bulk superconductivity. Topological properties of $Pb_2Pd$ are also confirmed by calculating Z2 invariants and surface states spectrum, which suggest the presence of strong topology in $Pb_2Pd$. This is the first report on calculation of topological invariants for $Pb_2Pd$. This report strongly represents $Pb_2Pd$ material as a possible material to realize topological superconductivity.

**Acknowledgment:**

The authors would like to thank Director NPL for his keen interest and encouragement. M.M. Sharma and N.K. Karn would like to thank CSIR, India for the research fellowship. M.M. Sharma, N.K. Karn are also thankful to AcSIR for Ph.D. registration. XPS measurement in CIF of Pondicherry University are also acknowledged.

**Conflict of Interest statement:**

Authors have no conflict of interest.

**Table-1**

Parameters obtained from Rietveld refinement:

| Cell Parameters | Refinement Parameters |
|---|---|
| Cell type: Tetragonal<br>Space Group: I4/mcm<br>Lattice parameters: a=b=6.863(5) Å & c=5.840(5) Å, $\alpha=\beta=\gamma=90°$<br>Cell volume: 275.068Å$^3$<br>Density: 12.577g/cm$^3$<br>Atomic coordinates: Pb (0.1643,0.6643,0)<br>Pd (0,0,0.25) | $\chi^2$=5.09<br>$R_p$=7.58<br>$R_{wp}$=9.68<br>$R_{exp}$=4.29 |

**Table-2**

XPS peaks position and FWHM of constituent elements of synthesized $Pb_2Pd$ single crystal:

| Element | Spin-orbit doublet | Binding Energy | FWHM |
|---|---|---|---|
| Pb | $4f_{7/2}$ | 138.55±0.005eV | 1.6±0.02eV |
| | $4f_{5/2}$ | 143.42±0.001eV | 1.6±0.01eV |
| Pd | $3d_{5/2}$ | 335.85±0.009eV | 0.7±0.08eV |
| | $3d_{3/2}$ | 341.15±0.012eV | 0.6±0.07eV |

**Table-3**

Parameters obtained from Heat capacity measurements:

| **Parameter** | **Obtained Value** |
|---|---|
| $T_c^{onset}$ | 2.86K |
| $\gamma_n$ | 5.72 ± 0.34 mJ mol$^{-1}$ K$^{-2}$ |
| $\beta_n$ | 3.18 ± 0.02 mJ mol$^{-1}$ K$^{-4}$ |
| $D_c(E_F)$ | 2.35 |
| $\theta_D$ | 122 ± 1K |
| $\lambda_{e\text{-}ph}$ | 0.70 |
| $\Delta C_{el}/\gamma_n T_c$ | 1.83 |
| $\alpha = 2\Delta(0)/k_B T_C$ | 5.68 |
| $\Delta(0)$ | 0.70meV |

**Figure Captions:**

**Fig. 1:** Schematic of heat treatment followed to synthesize $Pb_2Pd$ single crystal.

**Fig. 2:** Rietveld refined PXRD of synthesized $Pb_2Pd$ single crystal inset is showing the unit cell of the same drawn by using VESTA software.

**Fig. 3:** (a) SAED pattern of synthesized $Pb_2Pd$ crystal (b) HRTEM image of $Pb_2Pd$ crystal showing stacking of (002) planes. (c) SEM image of surface morphology of synthesized $Pb_2Pd$ crystal (d) EDX spectra of $Pb_2Pd$ crystal in which inset is showing the atomic percentage of the constitute elements.

**Fig. 4:** XPS spectra of synthesized $Pb_2Pd$ crystal in (a) Pb 4f region (b) Pd 3d region.

**Fig. 5:** AC magnetic susceptibility vs temperature plot of synthesized $Pb_2Pd$ single crystal.

**Fig. 6(a):** Zero Field C/T vs $T^2$ plot of synthesized $Pb_2Pd$ single crystal where solid black line represents a linear fit to $\frac{C}{T}=\gamma_n+\beta_n T$. The inset is showing a C/T vs T plot showing a $T_c^{onset}$ at 2.86K. **6(b):** Normalized electronic specific heat vs $T/T_c$ plot of synthesized $Pb_2Pd$ single crystal where the solid black line represents s-wave fitting of the same. **6(c):** Variation of C/T with temperature at different fields viz. 0Oe, 20Oe, 30Oe and 50Oe of $Pb_2Pd$ single crystal, the inset is showing variation of $T_c$ with applied field. **6(d):** $C_{el}/T$ vs T plot of synthesized $Pb_2Pd$ single crystal, the inset is showing the linear fitted plot of normalized Sommerfeld coefficient against the applied field.

**Fig. 7(a):** The first image shows the first Brillouin zone with high symmetric points, the green arrow shows the path chosen for the Band structure calculation. The other four images show the fermi surface of synthesized $Pb_2Pd$ crystal corresponding to the bands that are crossing the Fermi level. **7(b):** Calculated bulk electronic band structure along with the Density of States (DOS) w/o and with SOC within the protocols of Density Functional Theory (DFT). **7(c):** The zoomed view of the bands of the shaded region in bulk electronic band structure in Fig. 7(b).

**Fig. 8(a):** The (011) plane is shown with the path marked for surface spectrum calculation **8(b):** The image shows the gap energy dispersion contours in (011) plane, left shows without SOC and right one shows for with SOC case **8(c):** This shows the surface state spectrum, Dirac cone is observable at $\overline{\text{X}}$ point near energy -1.0eV. A similar Dirac cone of type-II is observable very close to fermi energy towards high symmetry point $\overline{\text{M}}$.

Fig. 1

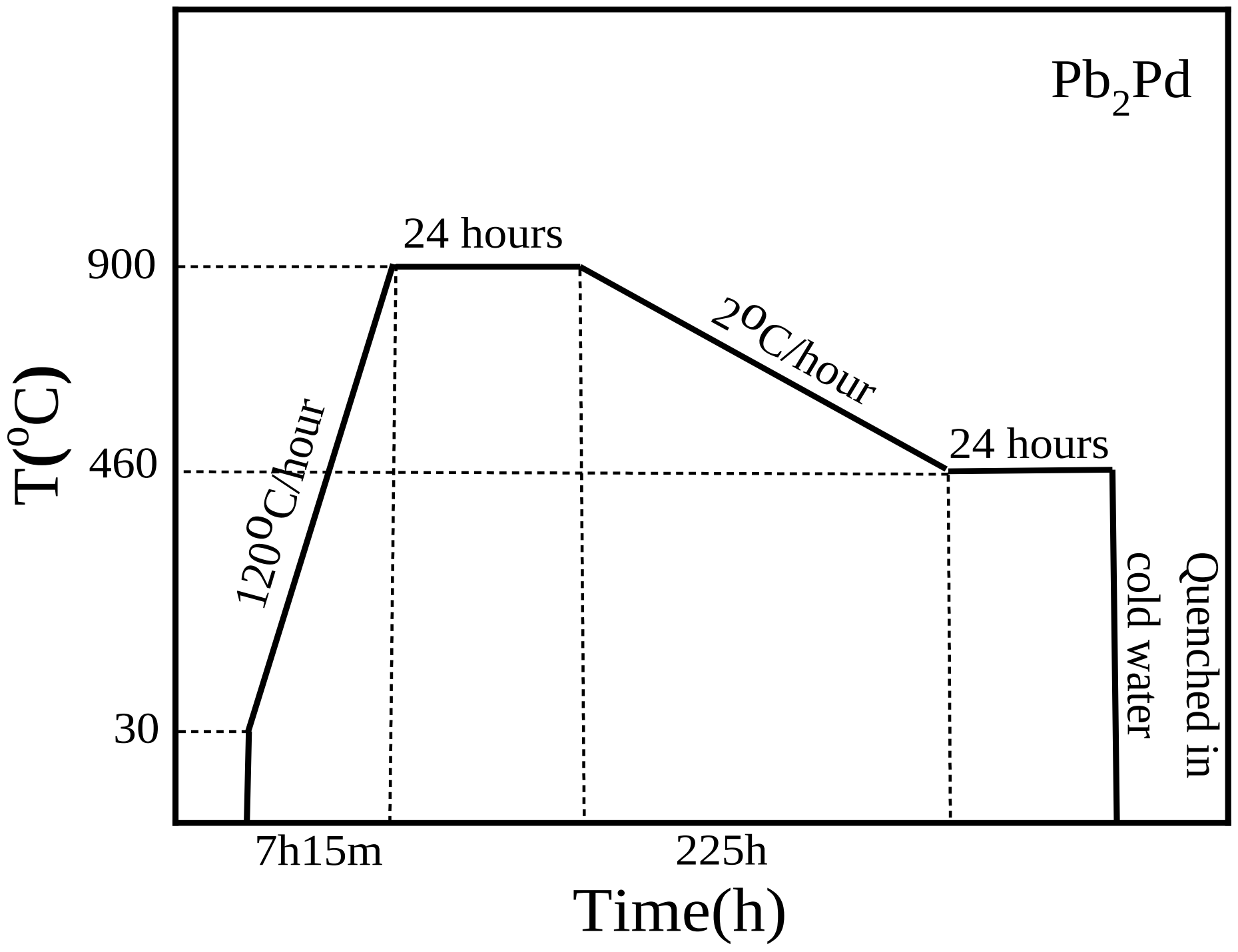

Fig. 2

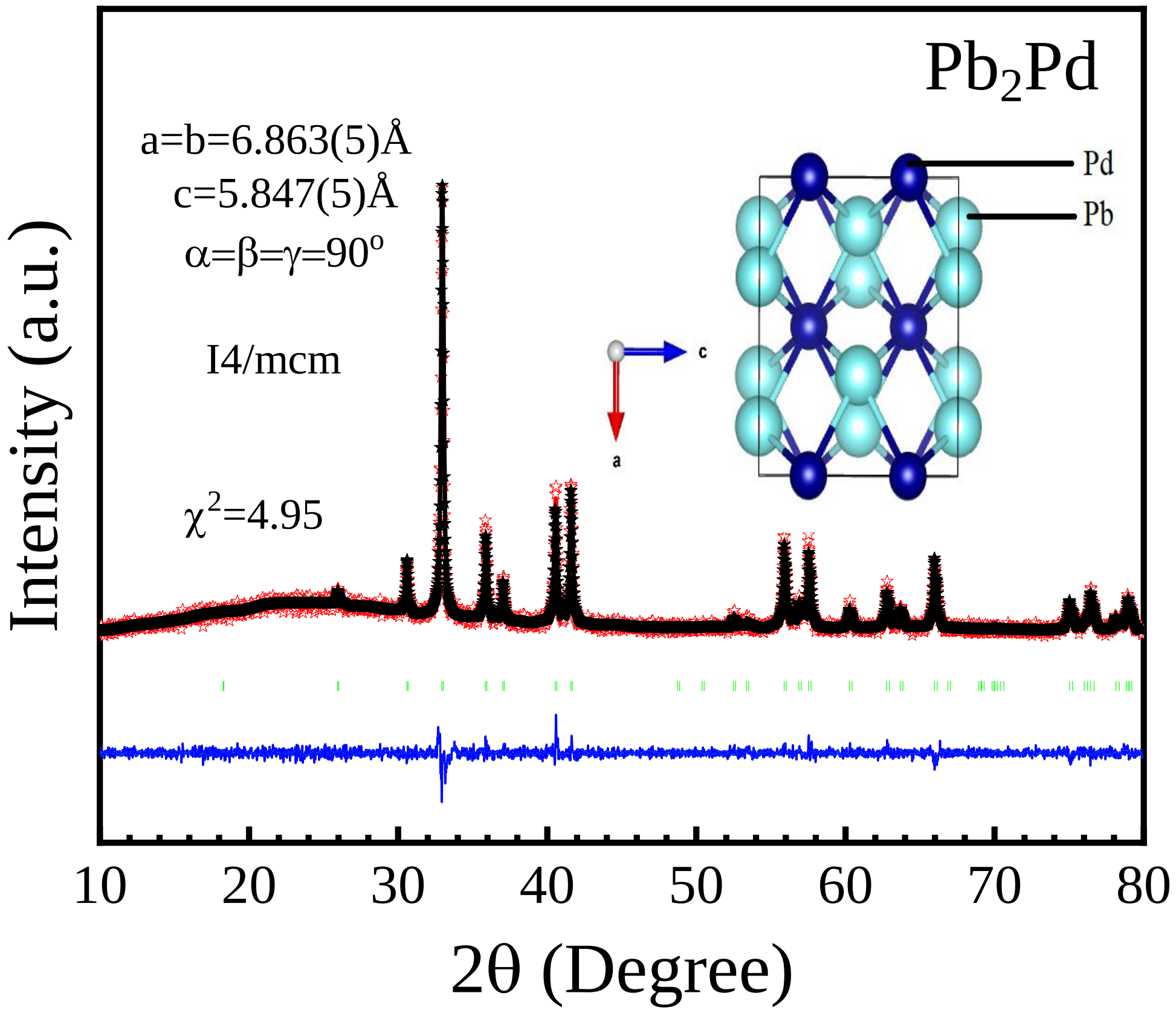

Fig. 3

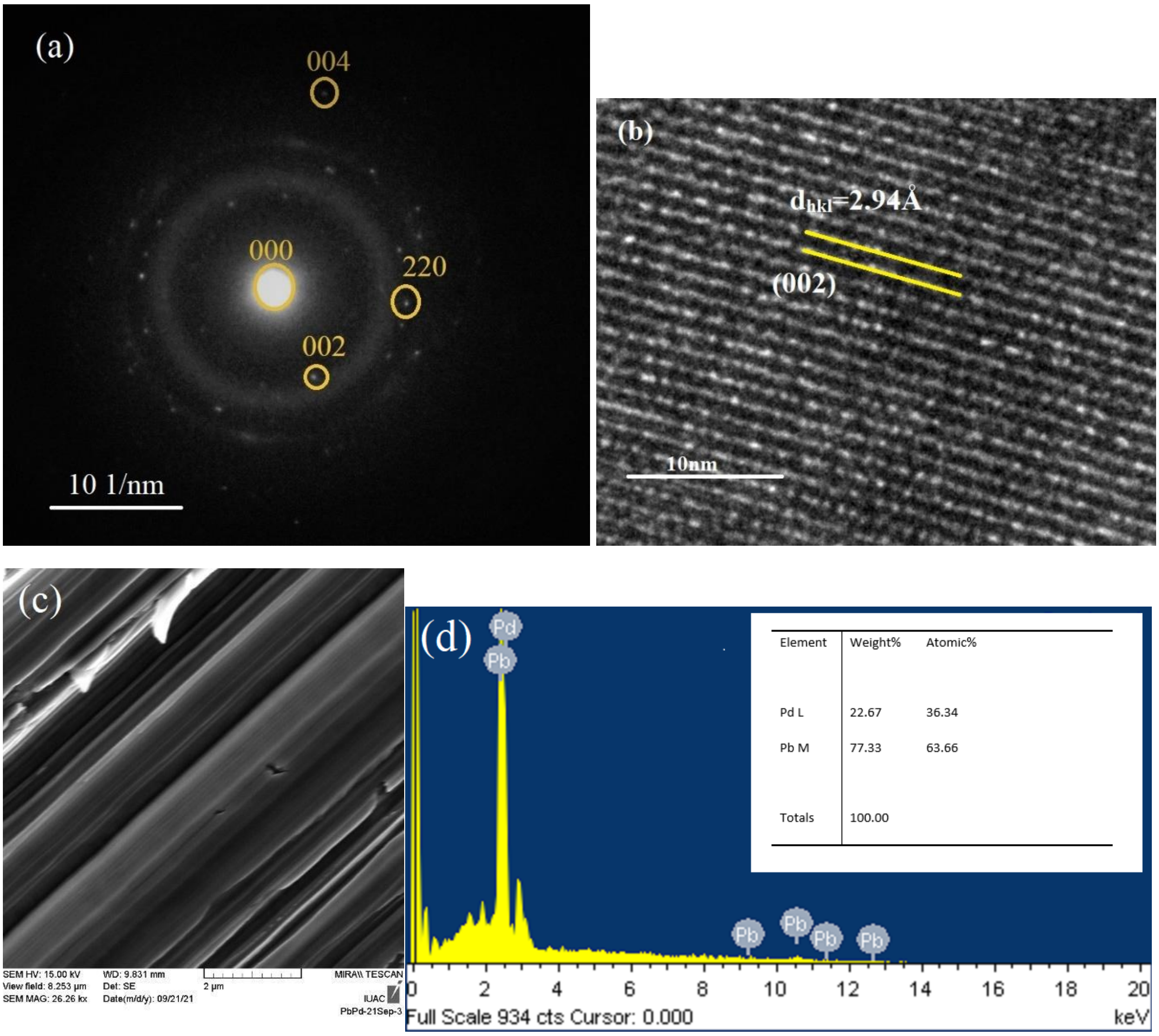
(a)
004
000
220
002
10 1/nm
(b)
dhkl=2.94Å
(002)
10nm
(c)
SEM HV: 15.00 kV
WD: 9.831 mm
View field: 8.253 µm
Det: SE
SEM MAG: 26.26 kx
Date(m/d/y): 09/21/21
2 µm
MIRA\\ TESCAN
IUAC
PbPd-21Sep-3
(d)
Pd
Pb
Element Weight% Atomic%
Pd L 22.67 36.34
Pb M 77.33 63.66
Totals 100.00
0 2 4 6 8 10 12 14 16 18 20
Full Scale 934 cts Cursor: 0.000
keV

Fig. 4

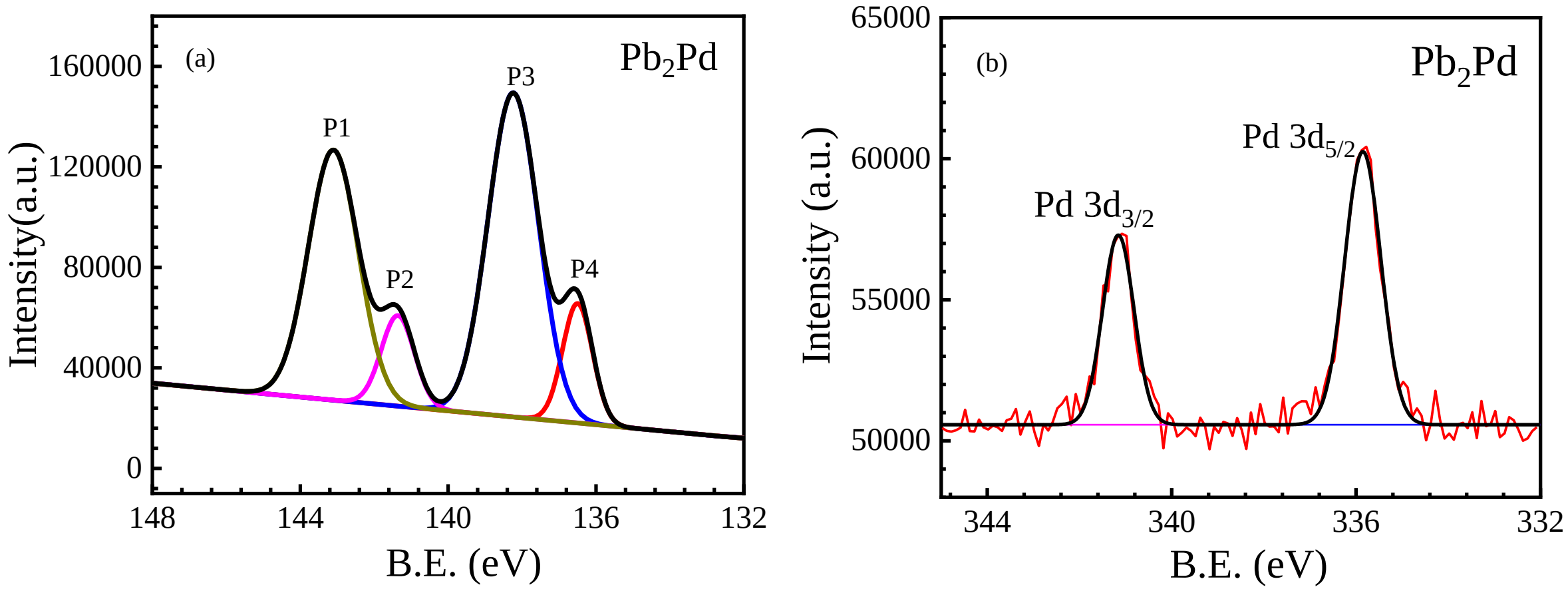

Fig. 5

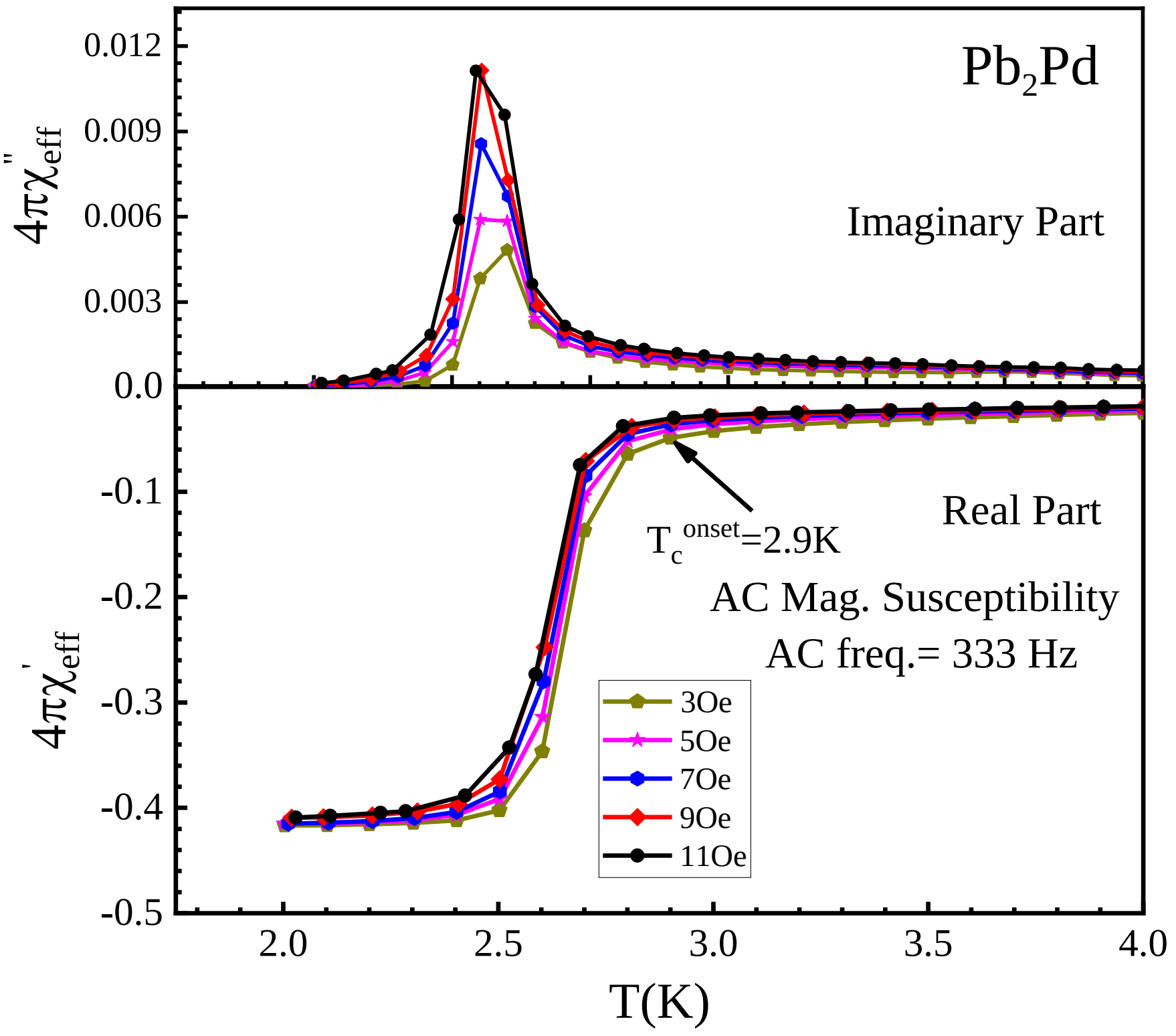

Fig. 6

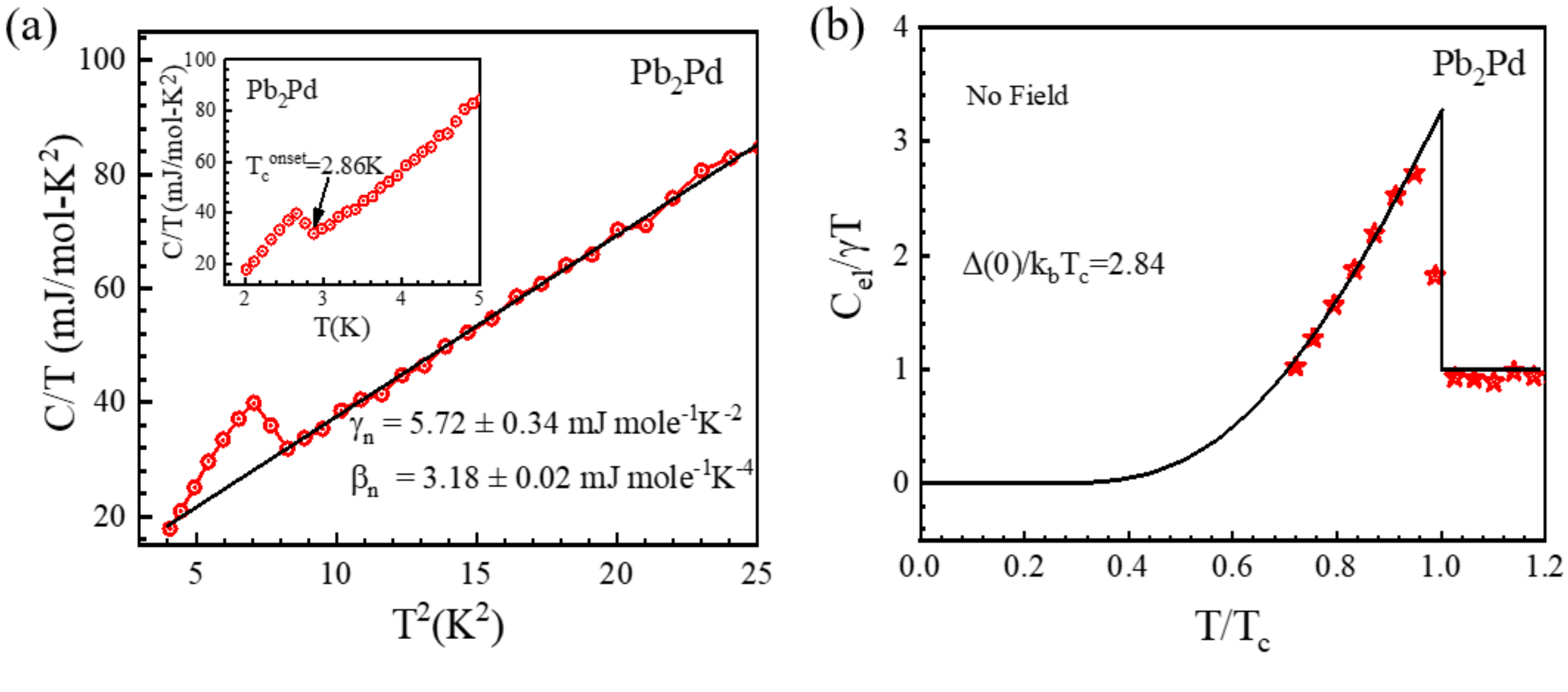

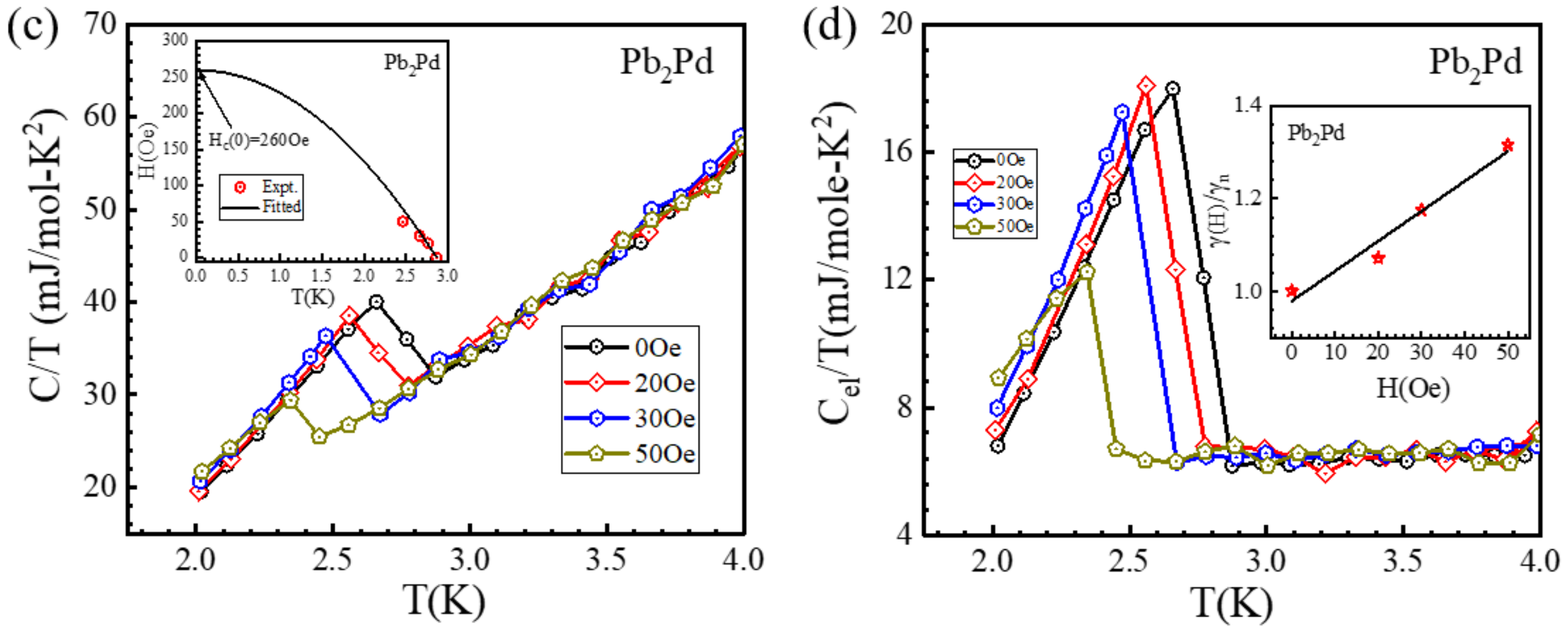

Fig. 7(a)

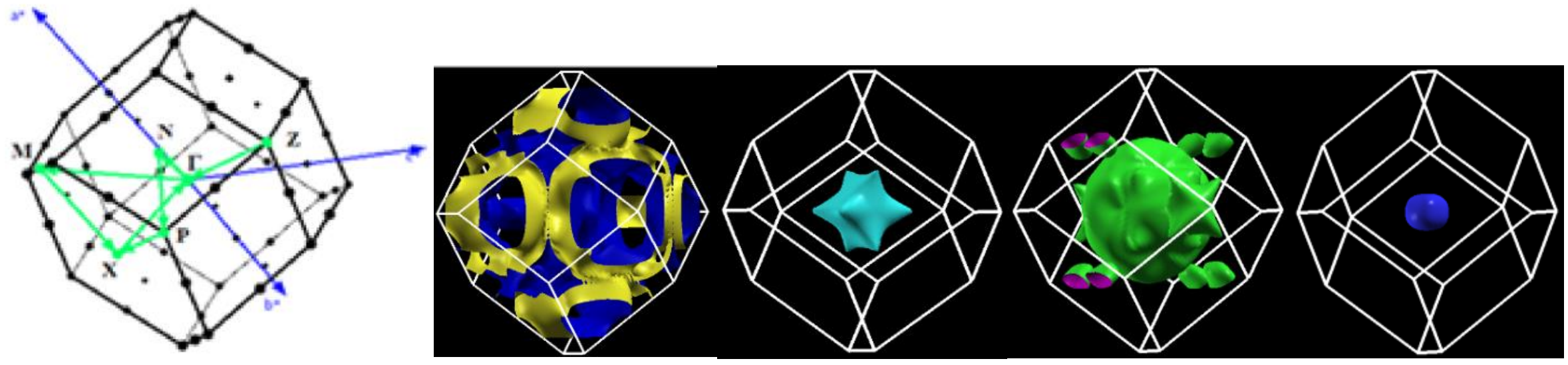

Fig. 7(b)

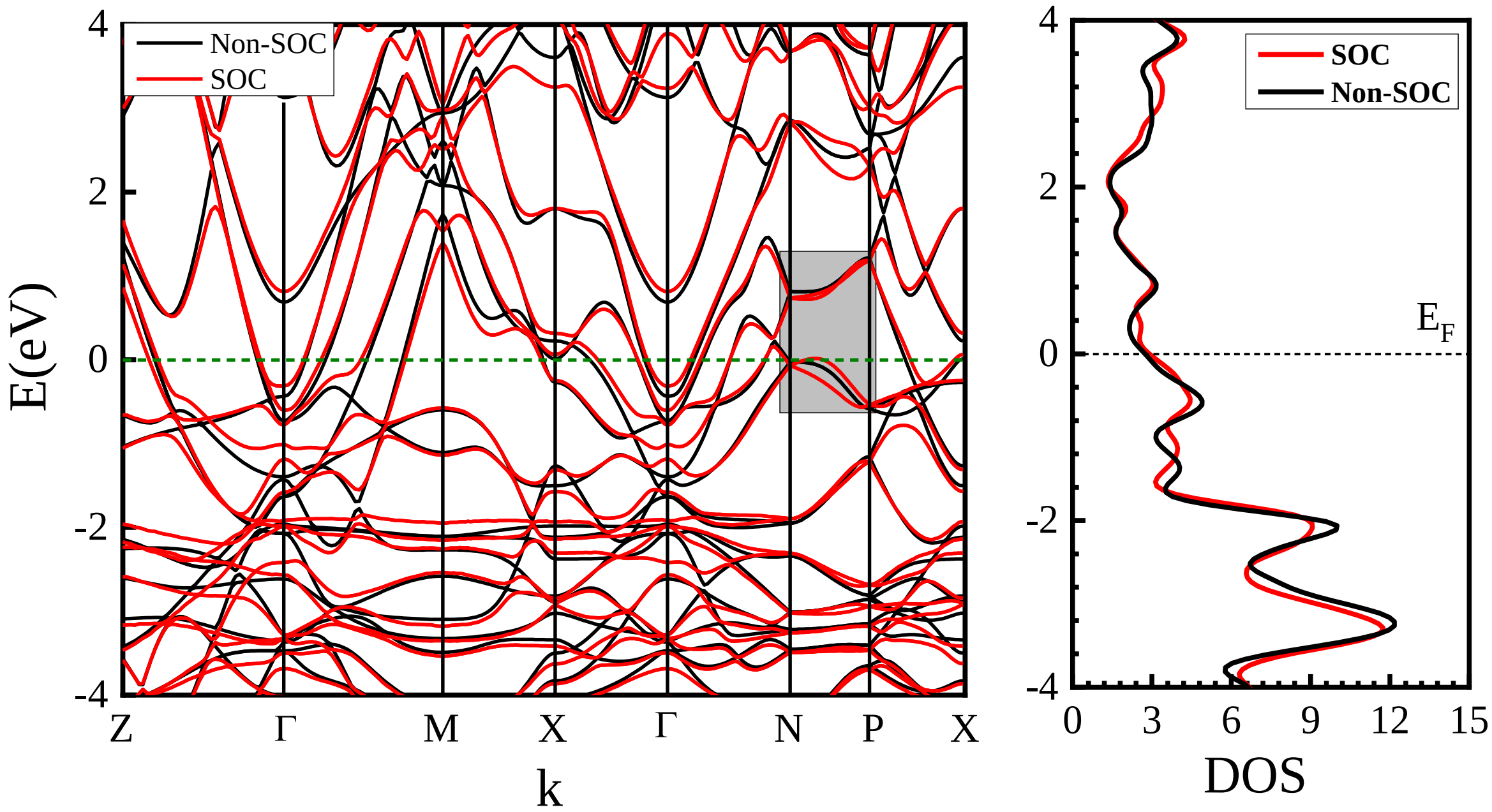

Fig. 7(c)

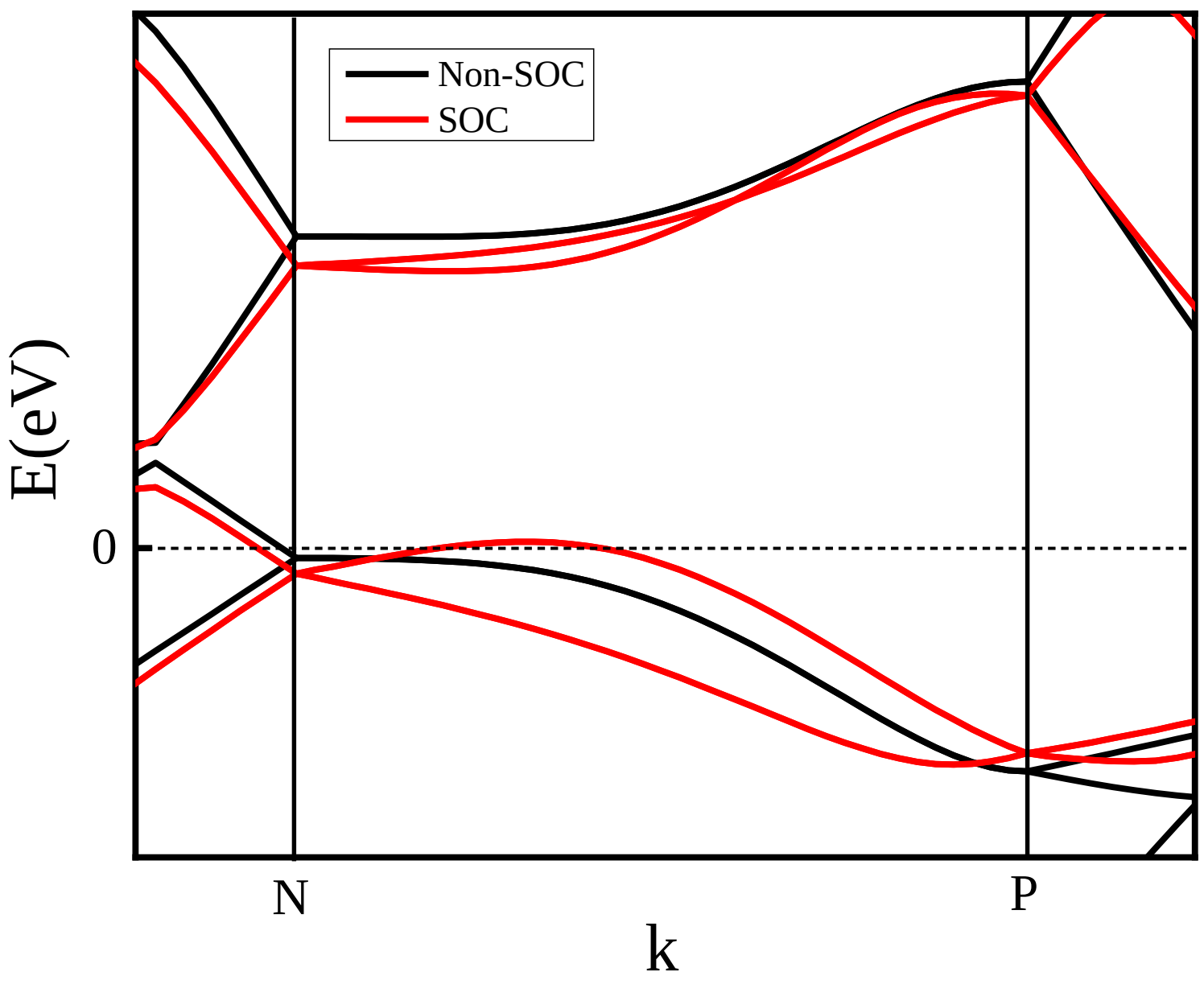

Fig. 8(a)

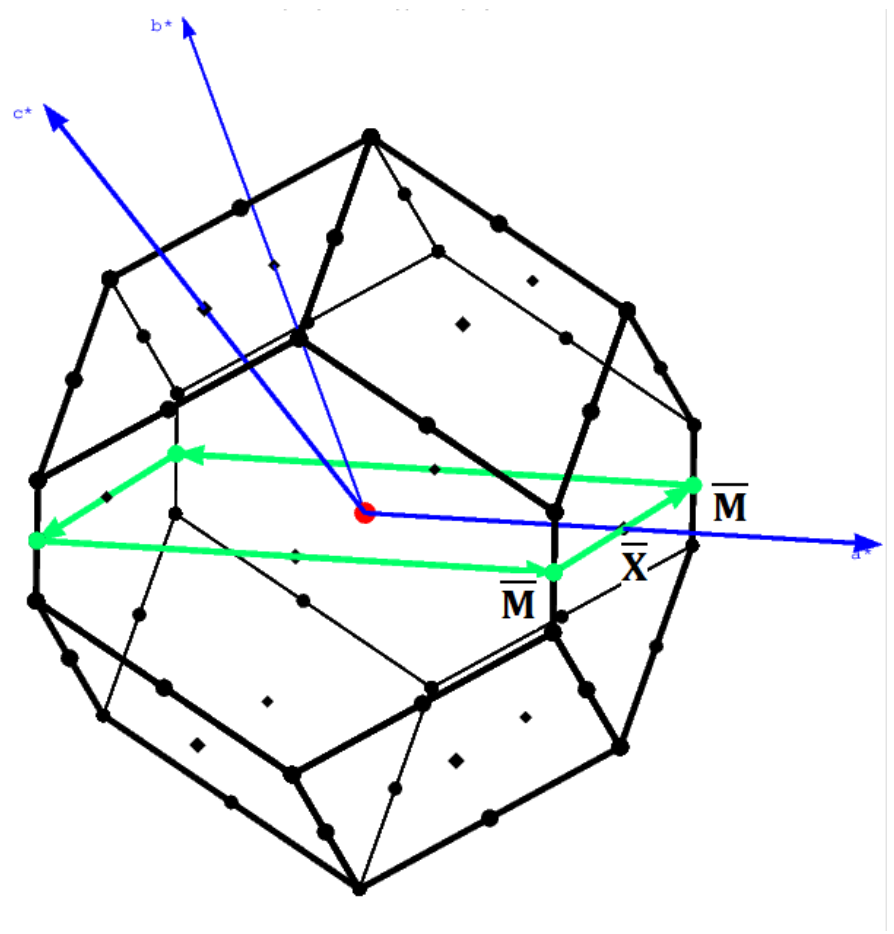

Fig. 8(b)

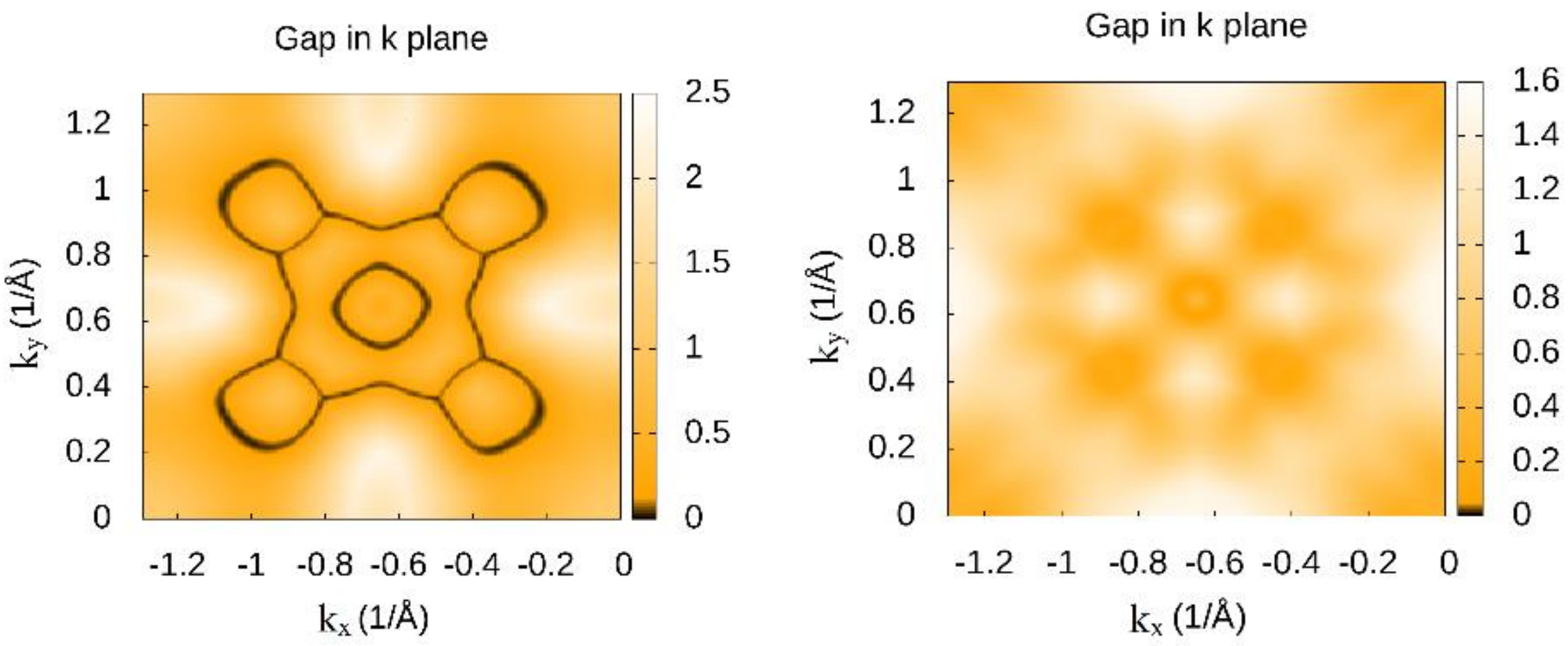

Fig. 8(c)

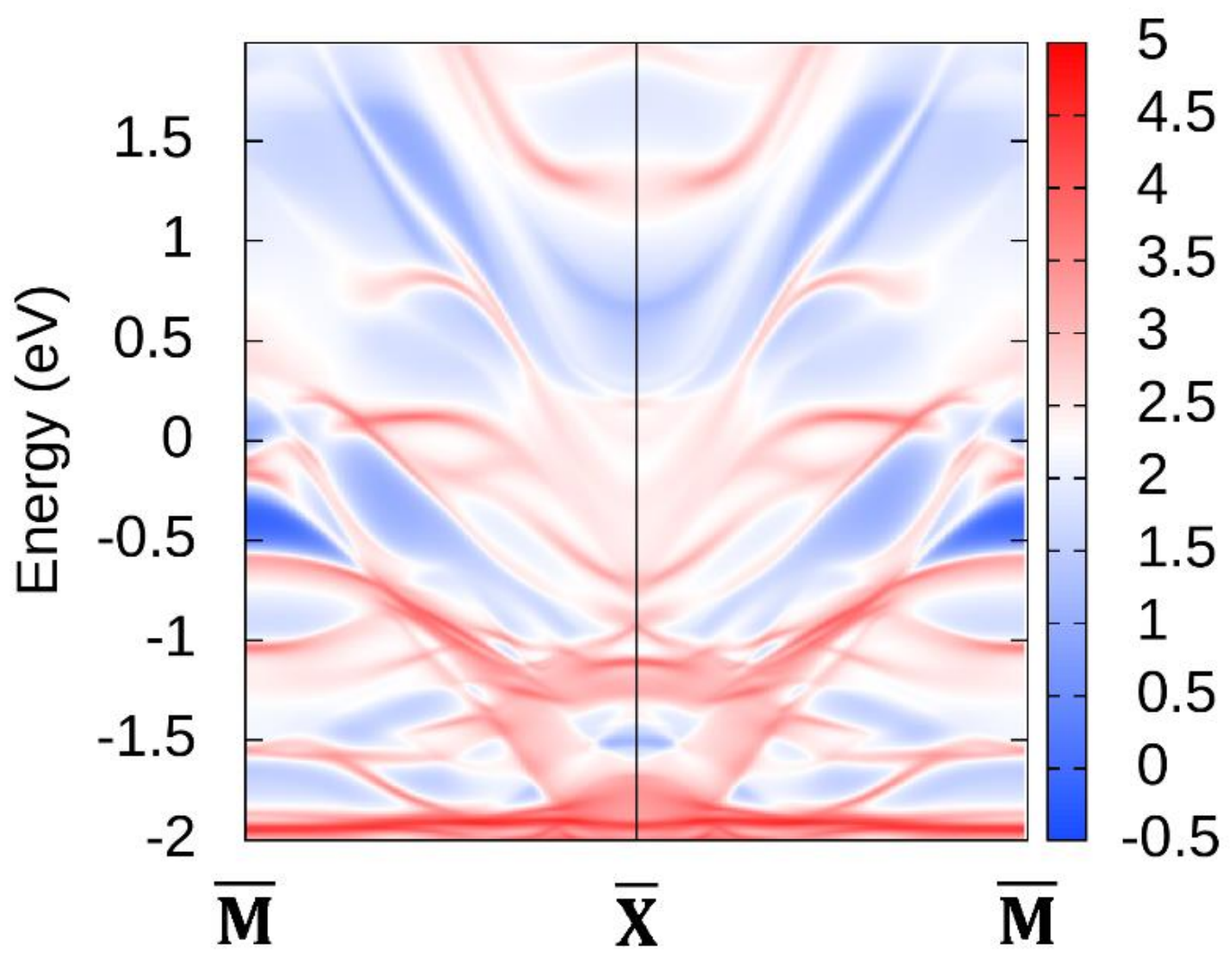